\documentclass[12pt,a4paper,twoside]{article}
\usepackage[T1]{fontenc}
\usepackage{amsmath,amsfonts,amssymb,amsthm,mathabx}
\usepackage{cite}
\usepackage{hyperref}
\usepackage{graphicx}
\usepackage{pdfsync}

\newcommand{\be}{\begin{equation}}
\newcommand{\ee}{\end{equation}}
\newcommand{\bea}{\begin{eqnarray}}
\newcommand{\eea}{\end{eqnarray}}

\def\d_Vphi{\mathrm{d}_V\hspace{-0.06em}\phi}
\def\d_Vphibar{\mathrm{d}_V\hspace{-0.06em}\bar\phi}
\def\d_Vxi{\mathrm{d}_V\hspace{-0.06em}\xi}

\def\be{\begin{eqnarray}}
\def\ee{\end{eqnarray}}
\def\beann{\begin{eqnarray*}}
\def\eeann{\end{eqnarray*}}
\def\beq{\begin{equation}}
\def\eeq{\end{equation}}
\def\ba{\begin{array}}
\def\ea{\end{array}}
\def\ben{\begin{enumerate}}
\def\een{\end{enumerate}}
\def\bea{\begin{eqnarray}}
\def\eea{\end{eqnarray}}

\def\5{\bar }
\def\6{\partial }
\def\7{\hat }
\def\4{\tilde }

\advance\textheight\topskip
\renewcommand{\tilde}{\widetilde}
\renewcommand{\hat}{\widehat}

\newcommand{\bref}[1]{\textbf{\ref{#1}}}

\newcommand{\p}[1]{|#1|}

\renewcommand{\p}{\partial}

\renewcommand{\geq}{\,{\geqslant}\,}
\renewcommand{\leq}{\,{\leqslant}\,}

\newcommand{\binner}[2]{%
  {\langle}\kern-4.15pt{\langle}#1{,}\,#2{\rangle}\kern-4.15pt{\rangle}}

\newcommand{\half}{\frac{1}{2}}

\newcommand{\ffrac}[2]{\raisebox{.5pt}%
  {\footnotesize$\displaystyle\frac{#1}{#2}$}\kern1pt}

\def\cL{\mathcal{L}}

\numberwithin{equation}{section} \makeatletter
\@addtoreset{equation}{section}

\DeclareFontFamily{OT1}{rsfs}{} \DeclareFontShape{OT1}{rsfs}{m}{n}{
<-7> rsfs5 <7-10> rsfs7 <10-> rsfs10}{}
\DeclareMathAlphabet{\mycal}{OT1}{rsfs}{m}{n}

\begin{document}

\title{Three-dimensional asymptotically flat Einstein-Maxwell theory}

\author{Glenn Barnich, Pierre-Henry Lambert and Pujian Mao}

\date{}

\def\mytitle{Three-dimensional asymptotically flat Einstein-Maxwell theory}

\pagestyle{myheadings} \markboth{\textsc{\small G.~Barnich,
    P.H.~Lambert, P.~Mao}}{%
  \textsc{\small 3d asymptotically flat Einstein-Maxwell theory}}

\addtolength{\headsep}{4pt}

\begin{centering}

  \vspace{1cm}

  \textbf{\Large{\mytitle}}

  \vspace{1.5cm}

  {\large Glenn Barnich, Pierre-Henry Lambert and Pujian Mao}

\vspace{.5cm}

\begin{minipage}{.9\textwidth}\small \it  \begin{center}
   Physique Th\'eorique et Math\'ematique \\ Universit\'e Libre de
   Bruxelles and International Solvay Institutes \\ Campus
   Plaine C.P. 231, B-1050 Bruxelles, Belgium
 \end{center}
\end{minipage}

\end{centering}

\vspace{1cm}

\begin{center}
  \begin{minipage}{.9\textwidth}
    \textsc{Abstract}.  Three-dimensional Einstein-Maxwell theory with
    non trivial asymptotics at null infinity is solved. The symmetry
    algebra is a Virasoro-Kac-Moody type algebra that extends the bms3
    algebra of the purely gravitational case. Solution space involves
    logarithms and provides a tractable example of a polyhomogeneous
    solution space. The associated surface charges are non-integrable
    and non-conserved due to the presence of electromagnetic news. As
    in the four-dimensional purely gravitational case, their algebra
    involves a field-dependent central charge.
  \end{minipage}
\end{center}

\vspace{1cm}
\thispagestyle{empty}
\newpage

\begin{small}
{\addtolength{\parskip}{-2pt}
 \tableofcontents}
\end{small}
\thispagestyle{empty}
\newpage

\section{Introduction}
\label{sec:introduction}

The original studies of four-dimensional asymptotically flat
spacetimes at null infinity
\cite{Bondi:1962px,Sachs1962a,Newman:1963yy} and their extensions to
include the electromagnetic field \cite{Burg1969,Exton1969} rely on an
expansion in inverse powers of the radial coordinate $r$ for the
metric components or the spin and tetrad coefficients. In order to
guarantee a self-consistent solution space, some of these expansions
need well-chosen gaps so as to prevent the appearance of logarithmic
terms in $r$.

In more recent investigations, this assumption has been relaxed.  More
general consistent solution spaces have been proposed that involve
double series with inverse powers and logarithms in $r$ from the very
beginning. Details on such ``polyhomogeneous spacetimes'' can be found
for instance
in~\cite{Winicour:1985pi,Chrusciel:1993hx,ValienteKroon:1998vn}.

Another non trivial aspect of 4d spacetimes with non trivial
asymptotics at Scri is that charges associated to the asymptotic
symmetry transformations, even though well-defined, are neither
conserved nor integrable \cite{Wald:1999wa}. Furthermore, when
considering a local version of the asymptotic symmetry algebra
\cite{Barnich:2009se,Barnich:2010eb}, the associated current algebra
acquires a field dependent central extension
\cite{Barnich:2011mi,Barnich:2013axa}.

In contrast, three-dimensional asymptotically flat Einstein gravity at
null infinity is much easier, in the sense that the expansion in
inverse powers of $r$ of the general solution with non trivial
asymptotics can be shown not to admit logarithms and to truncate after
the leading order terms \cite{Barnich:2010eb}. The symmetry algebra
\cite{Ashtekar1997} is $\mathfrak{bms}_3$, the charges are conserved,
integrable (and also $r$ independent \cite{Compere:2014cna}), while
their algebra involves a constant central extension
\cite{Barnich:2006avcorr}, closely related to the one for
asymptotically anti-de Sitter spacetimes \cite{Brown:1986nw}.

The purpose of the present paper is to study three-dimensional
Einstein-Maxwell theory with asymptotically flat boundary conditions
at null infinity. This model allows one to illustrate several aspects
of the four dimensional case in a simplified setting. On the one hand,
there is a clear physical reason for the occurrence of logarithms as
such a term is needed in the time component of the gauge potential in
order to generate electric charge. This term leads to a
self-consistent polyhomogeneous solution space that includes the
charged analog of particle \cite{Deser:1983tn} and cosmological
solutions
\cite{Ezawa:1992nk,Cornalba:2002fi,Cornalba:2003kd,Barnich:2012aw}. The
latter correspond to the flat space limit of the three-dimensional
charged rotating asymptotically anti-de Sitter black holes
\cite{Martinez:1999qi}. On the other hand, the asymptotic symmetry
algebra is a Virasoro-Kac-Moody type algebra that extends the
$\mathfrak{bms}_3$ algebra of the purely gravitational case. The
associated surface charges turn out to be neither conserved nor
integrable due to the presence of electromagnetic news. Furthermore
the algebra of surface charges now involves a field dependent central
charge that persists when switching off the news.

The plan of the paper is the following. In the next section, we work
out the asymptotic symmetry algebra. In section
\bref{sec:solution-space}, we present the polyhomogeneous solution
space, while section \bref{sec:charges} is devoted to the surface
charges and their algebra. In the last section, we show that upon
switching off the electromagnetic news, not surprisingly, charges
become conserved and integrable. Nevertheless, both the asymptotic
symmetry algebra and the central charge involve field dependent
terms. The appendix contains details on intermediate computations that
are omitted from the main text. Finally, a last section is devoted to
a comparison of the novel results derived here in three and previous
results obtained in four dimensions.

\section{Asymptotic symmetries}
\label{sec:asympt-symm}

To work out the asymptotic symmetries, we follow closely the original
literature \cite{Sachs1962} and adapt it to the current context. More
generally, for the Einstein-Yang-Mills system in all dimensions
greater than $3$, this problem has been addressed recently in detail
in a unified way both for flat and anti-de Sitter backgrounds in
\cite{Barnich:2013sxa}. In this approach, the gauge fixing condition
in the definition of asymptotic flat spacetimes fix the radial
dependence of gauge parameters completely, while the fall-off
conditions fix the temporal dependence. In the current set-up, the
fall-off conditions on $A_u$ are more relaxed as compared to those
considered in section 5.5 of \cite{Barnich:2013sxa} in order to
account for non-vanishing electric charge. As a consequence, the time
dependence of the electromagnetic gauge parameter is no longer fixed,
unless one switches off the news.

In order to define asymptotic flatness of the three-dimensional
Einstein-Maxwell at future null infinity, coordinates $u,r,\phi$ are
used together with the gauge fixing ansatz
\begin{equation}
\label{bms} {g_{\mu
    \nu}}=\begin{pmatrix} Ve^{2\beta}+r^2 U^2 & -e^{2\beta} & -r^2U\\
  -e^{2\beta}& 0 & 0 \\ -r^2U& 0 & r^2 \end{pmatrix},\quad A_r=0,
\end{equation}
where $U,\beta,V$ and $A_u,A_\phi$ are functions of $u,r,\phi$. Suitable
fall-off conditions that allow for non-vanishing electric charge are
\begin{equation}
\begin{split}
  \label{eq:1}
  U=o(r^{-1}),\quad V=o(r), \quad \beta=o(r^{0}),\\
  A_u=O(\ln{\frac{r}{r_0}}),\quad
  A_{\phi}=O(\ln{\frac{r}{r_0}}),
\end{split}
\end{equation}
where $r_0$ is a constant radial scale.

The gauge structure of Einstein-Maxwell theory can be described as
follows. Gauge parameters are pairs $(\xi^\mu,\epsilon)$ consisting of
a vector field $\xi^\mu\p_\mu$ and a scalar $\epsilon$. A generating
set of gauge symmetries can be chosen as
\begin{equation}
  \label{eq:5}
  -\delta_{(\xi,\epsilon)} g_{\mu\nu}=\cL_\xi
g_{\mu\nu},\quad -\delta_{(\xi,\epsilon)}A_\mu=\cL_\xi A_\mu
+\p_\mu\epsilon.
\end{equation}

When the gauge parameters are field
dependent, as will be the case for the parameters of asymptotic
symmetries below, the commutator of gauge transformations contains
additional terms:  
\begin{equation}
\left[\delta_{(\xi_1,\epsilon_1)},\delta_{(\xi_2,\epsilon_2)}\right](g_{\mu\nu},A_\mu)
=\delta_{\left[(\xi_1,\epsilon_1),(\xi_2,\epsilon_2)\right]_M}(g_{\mu\nu},A_\mu),
\end{equation}
where the Lie (algebroid) bracket for field dependent gauge parameters
is defined through
\begin{equation}
\begin{split}
 &  \left[(\xi_1,\epsilon_1),(\xi_2,\epsilon_2)\right]_M=(\hat\xi,\hat\epsilon),
    \\
 & \hat\xi=[\xi_1,\xi_2]
    +\delta_{(\xi_1,\epsilon_1)}\xi_2-
    \delta_{(\xi_2,\epsilon_2)}\xi_1\ ,\ \hat\epsilon=\xi_1(\epsilon_2)
    +\delta_{(\xi_1,\epsilon_1)} \epsilon_2-(1\leftrightarrow
    2).\label{eq:105}
\end{split}
\end{equation}

Gauge transformations preserving asymptotically flat configurations
are explicitly worked out in Appendix \bref{sec:asympt-symm-1}. They
are determined by gauge parameters depending linearly and
homogeneously on arbitrary functions $T(\phi),Y(\phi),E(u,\phi)$
according to
\begin{equation}
\label{symmetry}
\begin{split}
& \xi^u=f=T + u  Y',\\
& \xi^\phi=Y- f' \int_r^\infty \frac{e^{2\beta}}{r'^2}dr'
=Y - \frac{f'}{r}+o(r^{-2}),\\
& \xi^r=-r \p_\phi \xi^\phi +r U f'
=-r Y' + f'' + o(1), \\
& \epsilon=E(u,\phi) + f'
\int_r^\infty \frac{e^{2\beta} A_\phi}{r'^2}dr'=E(u,\phi)
+ O(\frac{\ln{\frac{r}{r_0}}}{r}),
\end{split}
\end{equation}
where dot and prime denote $u$ and $\phi$ derivatives, respectively.

Consider then the ``$\mathfrak{bms}_3$/Maxwell'' Lie algebra
consisting of triples $s=(T,Y,E)$ with bracket
\begin{equation}
  \label{eq:4}
  \left[s_1,s_2\right]=\left(\hat T,\hat Y,\hat E\right),
\end{equation}
where
\begin{equation}
  \label{eq:6}
  \hat T=Y_1T'_2+T_1Y'_2-(1\leftrightarrow
  2),\ \hat Y=Y_1Y'_2-(1\leftrightarrow
  2),\ \hat E=Y_1 E'_2+ f_1\dot E_2-(1\leftrightarrow
  2).
\end{equation}
This is the asymptotic symmetry algebra of the system in the following
sense:

{\em When equipped with the modified bracket \eqref{eq:105}, the
  parameters \eqref{symmetry} of the residual gauge symmetries form a
  representation of the Lie algebra \eqref{eq:4}. }

The proof, following the one originally worked out in
\cite{Barnich:2010eb}, is sketched in
Appendix~\bref{sec:asympt-symm-algebra-1}.

\section{Solution space}
\label{sec:solution-space}

In this section, we present the polyhomogeneous solution space for
our model, following mainly
\cite{Sachs1962a,Tamburino1966,Chrusciel:1993hx}. 

We start from the Einstein-Maxwell Lagrangian density in three
dimensions
\begin{equation}
\cL=\frac{\sqrt{-g}}{16\pi G}(R-F^2),\label{eq:2}
\end{equation}
with equations of motion
\begin{equation}
  \label{eq:7}
  \p_\nu (\sqrt{-g}F^{\mu\nu})=0, \quad L_{\mu\nu}:=G_{\mu\nu} - T_{\mu\nu}=0,
\end{equation}
where $T_{\mu\nu}= 2 F_{\mu\rho} F_\nu^{\;\;\rho} - \half g_{\mu\nu}
F^2$.

The detailed analysis in Appendix \bref{sec:solution-space-1} then
yields the following results: given the ansatz
\begin{equation}
  A_\phi=\alpha(u,\phi)\ln\frac{r}{r_0}+A_\phi^0(u,\phi)
+\sum\limits_{m=1}^\infty\sum\limits_{n=0}^{m}
\frac{A_{mn}(u,\phi)(\ln \frac{r}{r_0})^n}{r^m}, \label{eq:8}
\end{equation}
at a fixed time $u_0$, the general solution to the Einstein-Maxwell
system in three dimensions with the prescribed asymptotics is
completely determined in
terms of the initial data $A_\phi^0(u_0,\phi),A_{mn}(u_0,\phi)$, the
news functions $A_u^0(u,\phi)$ and integration functions
$\omega(\phi)$, $\lambda(\phi)$, $\theta(\phi)$, $\chi(\phi)$
according to 
\begin{eqnarray}
  \label{eq:9}
  && \alpha=-\omega-u \lambda',\quad N=\chi+u\theta'\\
  &&\beta=-\frac{\alpha^2}{2r^2}+\sum\limits_{m=1}^\infty\sum\limits_{n=0}^{m}
  \frac{\beta_{mn}(\ln \frac{r}{r_0})^n}{r^{m+2}},\\
  &&U=\frac{4\lambda \alpha \ln \frac{r}{r_0} + 2\lambda \alpha -N}{2r^2}
  +\sum\limits_{m=1}^\infty\sum\limits_{n=0}^{m} \frac{U_{mn}(\ln \frac{r}{r_0})^n}{r^{m+2}},\\
  &&A_u=-\lambda\ln {\frac{r}{r_0}}
  +A_u^0+\frac{\alpha'}{r}+\sum\limits_{m=1}^\infty
  \sum\limits_{n=0}^{m} \frac{{B}_{mn}(\ln \frac{r}{r_0})^n}{r^{m+1}},\\
  &&V=2 \lambda^2 \ln \frac{r}{r_0} + \theta +\frac{2\alpha \lambda' -
    2\lambda \alpha'}{r}
  +\sum\limits_{m=1}^\infty\sum\limits_{n=0}^{m}
  \frac{V_{mn}(\ln \frac{r}{r_0})^n}{r^{m+1}},
\end{eqnarray}
where the functions $\beta_{mn},U_{mn},B_{mn},V_{mn}$ are determined
recursively in terms of the initial data, the news, the integration
functions and their $\phi$ derivatives. In particular, 
\begin{equation}
\dot A_\phi^0
= -\lambda' + (A_u^0)'\label{eq:16}. 
\end{equation}
Furthermore, the leading parts of the metric and electromagnetic gauge
potentials are given by
\begin{eqnarray}
  &&ds^2=[2\lambda^2 \ln \frac{r}{r_0} + \theta +O(r^{-1})]du^2-[2+O(r^{-2})]du dr
\nonumber\\
  &&\hspace{1cm}-[4\lambda\alpha\ln \frac{r}{r_0} + 2\lambda\alpha-\chi-u
  \theta'+O(r^{-1}\ln
  \frac{r}{r_0})]du d\phi+ r^2 d\phi^2,\\
  &&A_\phi=\alpha \ln\frac{r}{r_0} + A_\phi^0+O(r^{-1}\ln
  \frac{r}{r_0}),\ A_u=-\lambda \ln\frac{r}{r_0} + A_u^0+O(r^{-1}).
\end{eqnarray}

Asymptotic symmetries transform solutions to solutions. This allows
one to work out the transformation properties of the functions
characterising asymptotic solution space: 
\begin{equation}
\begin{split}
  \cL_\xi g_{uu}:\ & -\delta \theta= Y \theta'+ 2 (\theta-\lambda^2)
  Y'- 2 Y''', \\
  & -\delta \lambda=Y \lambda'+ \lambda Y',\\
  \cL_\xi g_{u \phi}:\ & -\delta \chi=Y\chi'+ 2(\chi -2\omega \lambda)
  Y'+ T\theta'+ 2(\theta - \lambda^2)T' - 2T''',\\
  &-\delta \omega=Y \omega'+ \omega Y'+ T\lambda'+\lambda T',\\
  \cL_\xi A_u + \dot\epsilon:\ & -\delta A_u^0=Y
  (A_u^0)'+(A_u^0+\lambda) Y' + f
  \dot A_u^0  + \dot E,\\
  \cL_\xi A_\phi + \epsilon':\ & -\delta A_\phi^0=Y
  (A_\phi^0)'+(A_\phi^0-\alpha) Y'+ f \dot A_\phi^0 + A_u^0 f' +
  E'.\label{trafs}
\end{split}
\end{equation}

\section{Surface charge algebra}
\label{sec:charges}

Associating charges to asymptotic symmetries in general relativity is
a notoriously subtle question. The approach followed here consists in
deriving conserved co-dimension 2 forms in the linearized theory that
can be shown to be uniquely associated, up to standard amiguities, to
the exact symmetries of the background
\cite{Anderson:1996sc,Barnich:2001jy,Barnich:2003xg}. When using these
expressions for asymptotic symmetries in the full theory, neither conservation
nor integrability is guaranteed 
\cite{Wald:1999wa,Barnich:2007bf,Barnich:2011mi,Barnich:2013axa}.

More concretely, using the general expressions for the linearized
Einstein-Maxwell system derived in \cite{Barnich:2005kq}, the surface
charge one form of the linearized theory reduces to
\begin{equation}
  \oint_{S^\infty} \delta\hspace{-0.50em}\slash\hspace{-0.05em}k_{\xi,\varepsilon}
  =-\delta \oint_{S^\infty} K_{\xi,\varepsilon}+ \oint_{S^\infty}
K_{\delta\xi,\delta\varepsilon}-\oint_{S^\infty} \xi\cdot\Theta, \label{precharges}
\end{equation}
where
\begin{equation}
\begin{split}
  K_{\xi,\varepsilon}&=(dx^{n-2})_{\mu\nu}
\frac{\sqrt{-g}}{16\pi G}[\nabla^\mu \xi^\nu - \nabla^\nu \xi^\mu +
4 F^{\mu\nu}(\xi^\sigma A_\sigma +\epsilon)],\\
  \Theta&=(dx^{n-1})_{\mu}\frac{\sqrt{-g}}{16\pi G}
[\nabla_\sigma \delta g^{\mu\sigma}-\nabla^\mu \delta g^\nu_\nu + 4
F^{\sigma \mu}\delta A_\sigma],
\end{split}
\end{equation}
and $S^\infty$ is the circle at constant $u=u_0$ and $r=R\to
\infty$. At this stage, we have used already that expressions in
$\oint_{S^\infty}
\delta\hspace{-0.50em}\slash\hspace{-0.05em}k_{\xi,\varepsilon}$ that
are proportional to the exact generalized Killing equations vanish,
\begin{equation}
\begin{split}
  &  \frac{1}{16\pi G}\oint_{S^\infty}  \delta g^\nu_\rho
  (\nabla^\mu \xi^\rho+\nabla^\rho \xi^\mu)\sqrt{-g}(d^{n-2}x)_{\mu\nu} =0,\\
  &  \frac{1}{4\pi G}\oint_{S^\infty} g^{\mu\rho} \delta A^\nu
(\cL_\xi A
_\rho + \p_\rho \varepsilon)\sqrt{-g}(d^{n-2}x)_{\mu\nu}=0,
\end{split}
\end{equation}
when evaluated for solutions and asymptotic symmetry parameters
\eqref{symmetry}. As in four-dimensional asymptotically flat pure
Einstein gravity \cite{Wald:1999wa}, the remaining expression then
splits into an integrable part and a non-integrable part proportional
to the electromagnetic news,
\begin{equation}
  \label{eq:13}
  \oint_{S^\infty}
\delta\hspace{-0.50em}\slash\hspace{-0.05em}k_{\xi,\varepsilon}=
\delta Q_{s}+\Theta_s,
\end{equation}
with
\begin{equation}
\begin{split}
  & Q_s[g,A] =\frac{1}{16\pi G} \int_0^{2\pi} d\phi \Big[\theta T+Y
  (\chi+4 \lambda A_\phi^0) + 4 \lambda E\Big],
  \\
  & \Theta_{s}[\delta g,\delta A;g,A] =\frac{1}{4\pi G} \int_0^{2\pi}
  d\phi f A_u^0 \delta \lambda.\label{charges}
\end{split}
\end{equation}
It follows that $(8G)^{-1}\theta,(8G)^{-1}(\chi+4\lambda
A^0_\phi),(2G)^{-1}\lambda$ can be interpreted as the mass, angular
momentum and electric charge aspect, respectively. 

Applying now the proposal of \cite{Barnich:2011mi,Barnich:2013axa} for
the modified bracket of the integrable part of the charges,
\begin{equation}
\{Q_{s_1},Q_{s_2}\}=-\delta_{s_2} Q_{s_1} +
\Theta_{s_2}[-\delta_{s_1}g,-\delta_{s_1}A;g,A],
\end{equation}
gives
\begin{equation}
  \{Q_{s_1},Q_{s_2}\}=Q_{[s_1,s_2]} + K_{s_1,s_2},
\end{equation}
where
\begin{equation}
  K_{s_1,s_2}=\frac{1}{8 \pi G} \int_0^{2\pi}d\phi\ \left[Y_1' T_2''-
    2\lambda f_1 \dot E_2-\lambda^2 T_1 Y_2'- (1\leftrightarrow 2)\right].
\end{equation}
It is then straightforward to check that the field dependent central
extension satisfies the generalised cocycle condition
\begin{equation}
K_{[s_1,s_2],s_3}-\delta_{s_3} K_{s_1,s_2} +{\rm cyclic}\ (1,2,3)=0.
\end{equation}

\section{Switching off the news}
\label{sec:switching-news}

In the analysis above, we are in an unusual situation where the
asymptotic symmetry algebra depends arbitrarily on time through the
dependence of $E$ on $u$. This can be fixed by requiring the
electromagnetic news function to vanish, $A_u^0=0$. Since asymptotic
symmetries need to preserve this condition, we find from \eqref{trafs}
that $E=\bar E-\int_{u_0}^udu'\lambda Y'$. The asymptotic symmetry
algebra then becomes time independent, but field dependent since the
last of \eqref{eq:6} gets replaced by
\begin{equation}
\hat{\bar E}=Y_1 \bar E'_2-
T_1Y'_2\lambda-(1\leftrightarrow 2).\label{eq:14}
\end{equation}
Charges become integrable and conserved: the second, non integrable
part vanishes while in the first line of \eqref{charges}, $E,A^0_\phi$
get replaced by $\bar E,\bar A^0_\phi$. In order to see this, one has
to go back to \eqref{precharges} where the second term now contributes to
remove the $u$-dependent terms when using that, on shell,
$E=\bar E(\phi)-u\lambda(\phi)Y'$, and
$A^0_\phi=\bar A^0_\phi(\phi) -u\lambda'$. Finally, the field
dependent central charge becomes
\begin{equation}
  \label{eq:15}
  K_{s_1,s_2}=\frac{1}{8 \pi G} \int_0^{2\pi}d\phi\ \left[Y_1' T_2''
    +\lambda^2 T_1 Y_2'- (1\leftrightarrow 2)\right].
\end{equation}

\section{Discussion}
\label{sec:conclusion}

Apart from its intrinsic interest, one might hope that the elaborate
symmetry structure and the explicit solution of the three-dimensional
Einstein-Maxwell system with non trivial asymptotics at Scri presented
here could be suitably tuned so as to have applications in the context
of holographic condensed matter models in 1+1 dimensions. Indeed, the
Einstein-Maxwell system with various backgrounds, asymptotics, and
additional scalar or form fields is ubiquitous in this context, see
for instance
\cite{Hartnoll:2008vx,Kachru:2008yh,Balasubramanian:2008dm}, and more
specifically \cite{Maity:2009zz,Ren:2010ha,Jensen:2010em} in three
bulk dimensions. From the viewpoint of symmetries as well, this is
quite reasonable since the $\mathfrak{bms}_3$ algebra is isomorphic to
$\mathfrak{gca}_2$, the Galilean conformal algebra in $2$ dimensions
\cite{Bagchi:2009pe,Bagchi:2010zz}.

Independently of such speculations, let us compare the
three-dimensional results derived here to those of the four
dimensional case. First, we note that in the four dimensional
Einstein-Maxwell system, one imposes the conditions $A_r=0$ and
$A_u=O(r^{-1})$ (see e.g.~\cite{Burg1969} and \cite{Ashtekar:1987tt}
section II.C for a detailed discussion of pure electromagnetism). As
shown in \cite{Barnich:2013sxa}, from the viewpoint of asymptotic
symmetries, the absence of a term in $A_u$ of order zero in $r^{-1}$
also guarantees a time independent symmetry algebra similar to the one
discussed here in three dimensions, but with an additional arbitrary
dependence on the supplementary polar angle. In four dimensions, the
electromagnetic news nevertheless persists since it is encoded in
different components of the vector potential.

Concerning the algebra of charges, there does not exist, to our
knowledge, a complete study of the Einstein-Maxwell system in the
four-dimensional case. That is the reason why we compare the rest of
the results here to the purely gravitational ones in four dimensions.

As recalled in the introduction, in four dimensions, self-consistent
asymptotically flat solution spaces at Scri including charged black
hole solutions have been constructed in spaces involving integer
powers of $1/r$. 

The simplest solution in three dimensions is the flat limit of the
charged BTZ black hole. It is characterized by
$\omega=0=A^0_\phi=A_{mn}=A^0_u$ and $\theta=8GM$, $\chi=8G J$,
$\lambda=2G Q$, where $M,J,Q$ are constants that, according to
\eqref{charges}, are interpreted as the mass, angular momentum and
electric charge of the solution. Let us recall that the uncharged
solutions with $Q=0$ describe three dimensional cosmologies
\cite{Cornalba:2002fi,Cornalba:2003kd,Barnich:2012aw} when $M\geq 0$
and spinning particles, i.e., the angular defects and excesses of
\cite{Deser:1983tn}, when $M <0$.

In both cases, it is the news, electromagnetic in the former and
gravitational in the latter, that is responsible for the
non-integrability and non conservation of the charges. In the latter
case, there is no (field-dependent) central extension, unless one
admits singular symmetry generators at null infinity and considers a
local extension of the $\mathfrak{bms}_4$ algebra including
superrotations
\cite{Barnich:2010eb,Barnich:2011mi,Barnich:2013axa}. It disappears
when switching off the news, as this reduces super to standard Lorentz
rotations. In the former case, superrotations always exist and are
globally well-defined at null infinity. The field dependent central
extension persists even after switching off the electromagnetic
news. To our knowledge, this is the first example in the context of
asymptotic symmetries where there is a field dependent term in the
symmetry algebra and in the central extension of the algebra of
conserved and integrable charges.

The first field independent term in \eqref{eq:15} exists also for pure
gravity in three dimensions and is well understood from a
cohomological point of view
\cite{Barnich:2006avcorr,Barnich:2011ct}. It has also been used in an
argument pertaining to the Bekenstein-Hawking entropy of the
three-dimensional cosmological solutions
\cite{Barnich:2012xq,Bagchi:2012xr}, modeled on the one in
\cite{Strominger:1998eq} for the BTZ black holes.

The second field dependent term involving the electric charge aspect,
is novel and much less understood. It certainly deserves further
study, both from the viewpoint of Lie algebroid cohomology and from a
physical perspective.

Finally, let us comment on the nature of the solutions considered
here. As in the majority of the papers on the subject since the
pioneering work by Bondi et al.~ \cite{Bondi:1962px}, the solutions
are constructed as formal power series in the radial coordinate. In
the polyhomogeneous case, there has been an investigation of
convergence and existence of such solutions for linear massless higher
spin fields on Minkowski spacetime as a preliminary study for the
gravitational problem \cite{ValienteKroon:2000jy}. Addressing this
question is clearly relevant in this set-up as well, but beyond the
scope of the current work. We just note that the asymptotic symmetry
algebra itself is not very sensitive to the details of solution space,
as it is based solely on (\ref{bms}), (\ref{eq:1}) and the absence of
news in later considerations.

\section*{Acknowledgements}

\addcontentsline{toc}{section}{Acknowledgments}

This work is supported in part by the Fund for Scientific
Research-FNRS (Belgium), by IISN-Belgium, by ``Communaut\'e fran\c
caise de Belgique - Actions de Recherche Concert\'ees'', and by the
Kavli Institute for Theoretical Physics China at the Chinese Academy
of Sciences through the program ``Quantum Gravity, Black Holes and
Strings''. P.~Mao is supported by a PhD fellowship from the China
Scholarship Council. He thanks Hernan A. Gonzalez and Hong-Bao Zhang
for useful discussions. P.-H.~Lambert is grateful to Laura Donnay for
useful discussions.

\appendix

\section{Details on computations}
\label{sec:details-computations}

\subsection{Residual symmetries}
\label{sec:asympt-symm-1}

Gauge parameters $\xi^\mu$ that preserve the metric ansatz depend on
two arbitrary functions $T(\phi),Y(\phi)$:
\begin{itemize}
\item $\cL_\xi g_{rr}=0$ implies $\partial_r\xi^u=0$ and so
  $\xi^u=f(u,\phi)$,
\item $\cL_\xi g_{r\phi}=0$ implies
  $\partial_r\xi^\phi=\dfrac{e^{2\beta}}{r^2}\partial_\phi f$ and so
  $\xi^\phi=Y(u,\phi)-\int_r^\infty
  dr'\dfrac{e^{2\beta}}{{r'}^2}\partial_\phi f$,
\item $\cL_\xi g_{\phi\phi}=0$ implies
  $\xi^r=-r(\partial_\phi\xi^\phi-U\partial_\phi f)$,
\item $\cL_\xi g_{u\phi}=o(r)$ implies $\partial_u Y=0$ and so
  $Y=Y(\phi)$,
\item $\cL_\xi g_{ur}=o(r^{0})$ implies $\partial_u f=\partial_\phi
  Y$ and so $f=T(\phi)+u\partial_\phi Y$,
\item $\cL_\xi g_{uu}=o(r)$ implies no further conditions.
\end{itemize}
The gauge parameter $\epsilon$ preserving the gauge and fall-off
conditions of the gauge potentials depends on an arbitrary function
$E(u,\phi)$ according to
\begin{itemize}
\item $\cL_\xi A_r+\p_r \varepsilon=0$ implies
  $\varepsilon=E(u,\phi) + \p_\phi \xi^u \int_r^\infty
  \frac{e^{2\beta}A_\phi}{r'^2}dr'$,
\item $\cL_\xi A_u+\p_u \varepsilon=O(\ln{\frac{r}{r_0}})=\cL_\xi
  A_\phi+\p_\phi \varepsilon$ imply no further conditions.
\end{itemize}

\subsection{Asymptotic symmetry algebra}
\label{sec:asympt-symm-algebra-1}

We want to show that the gauge parameters \eqref{symmetry}, when
equipped with the bracket \eqref{eq:105}, provide a representation of
the Lie algebra \eqref{eq:4}. By evaluating $\cL_{\xi}g_{\mu\nu}$, we
find
\begin{gather}
  \left\{\begin{array}{l}\label{fields}
      -\delta \beta =\xi^\alpha\p_\alpha\beta+\half
\big[\p_u f+\p_r\xi^r+\p_\phi f U],\\
      -\delta U =\xi^\alpha \partial_\alpha U + U \big[\p_u
      f+ \partial_\phi f U-\p_\phi \xi^\phi\big]- \partial_u
      \xi^\phi-\p_r \xi^\phi V + \partial_\phi
      \xi^r\frac{e^{2\beta}}{r^2},
\end{array}
\right.
\end{gather}
while $ -\delta A_\phi=\xi^\alpha\p_\alpha A_\phi + A_\alpha \p_\phi
\xi^\alpha + \p_\phi \epsilon$.  It follows that
\begin{gather}
  \left\{\begin{array}{l}\label{parameters}
      \delta_1 \xi^u_2=0,\\
      \delta_1 (\p_r\xi^\phi_2)=\p_\phi f_2 \frac{e^{2\beta}}{r^2} 2\delta_1 \beta,\\
      \delta_1 \xi^r_2=-r\big[\p_\phi(\delta_1\xi^\phi_2)-\p_\phi f_2 \delta_1 U\big],\\
      \delta_1 (\p_r \epsilon_2)=-\frac{1}{r^2}\left(\p_\phi f_2
        e^{2\beta} \delta_1 A_\phi + \p_\phi f_2
        e^{2\beta}A_\phi2\delta_1 \beta \right).
\end{array}
\right.
\end{gather}
Direct computation then shows that
\begin{eqnarray*}
  &&\p_r \hat \xi^u=\p_r \hat{f}=0,\hspace{1cm}\p_u \hat{f}=\p_\phi
  \hat{Y}, \hspace{1cm}\hat{f}=\hat{T}+u\p_\phi\hat{Y}, \\
  &&\p_r
  \hat \xi^\phi=\frac{e^{2\beta}\hat{f}}{r^2},\hspace{1cm}
\lim_{r\to\infty}\hat \xi^\phi=\hat{Y},  \\
  &&\hat \xi^r=-\p_\phi \hat \xi^\phi+ U \p_\phi \hat{f},\\
  &&\p_r \hat{\epsilon}=-\frac{\p_\phi \hat{f} e^{2\beta}
    A_\phi}{r^2},\hspace{1cm}\lim_{r\to\infty}\hat{\epsilon}=\hat{E},
\end{eqnarray*}
which proves the result since these conditions determine uniquely
gauge parameters \eqref{symmetry} where $(T,Y,E)$ have been replaced
by $(\hat T,\hat Y,\hat E)$.

\subsection{Solution space}
\label{sec:solution-space-1}

The equations of motion can be organized as as follows
\begin{eqnarray}
&&\p_\nu (\sqrt{-g}F^{u\nu})=0,\label{main1} \\
&&\p_\nu (\sqrt{-g}F^{\phi\nu})=0,\label{main2} \\
&&\p_\nu (\sqrt{-g}F^{r\nu})=0,\label{supp1} \\
&&L_{r\alpha}=G_{r\alpha} - T_{r\alpha}=0,\label{main3} \\
&&L_{\phi\phi}=G_{\phi\phi} - T_{\phi\phi}=0,\label{trivial1} \\
&&L_{u\phi}=G_{u\phi} - T_{u\phi}=0,\label{supp2} \\
&&L_{u u}=G_{u u} - T_{u u}=0. \label{supp3}\label{eq:3}
\end{eqnarray}
When equations (\ref{main1}) and (\ref{main2}) hold, the
electromagnetic Bianchi equation reduces to
$\p_r [\p_\nu (\sqrt{-g}F^{r\nu})]=0$. This means that if
$\p_\nu (\sqrt{-g}F^{r\nu})=0$ for some constant $r$, it vanishes for
all $r$. The gravitational Bianchi identities can be written as
\begin{equation}
0=2\sqrt{-g}\nabla_\nu
G_\mu^\nu=2 \p_\nu (\sqrt{-g}L_\mu^\nu) + \sqrt{-g} L_{\rho\sigma}
\p_\mu g^{\rho\sigma} +2\sqrt{-g}\nabla_\nu T_\mu^\nu.\label{eq:11}
\end{equation}
When (\ref{main1})-(\ref{main3}) are satisfied and $\mu=r$ in
\eqref{eq:11}, one gets $L_{\phi\phi} \p_r g^{\phi\phi}=0$ which
implies $L_{\phi\phi}=0$. In this case, the remaining Bianchi
identities reduce to
$2 \p_\nu (\sqrt{-g}L_\phi^\nu)=0=2 \p_\nu (\sqrt{-g}L_u^\nu)$. The
first one gives $\p_r(r L_{u\phi})=0$. This means that if
$r L_{u\phi}=0$ for some fixed $r$, it vanishes everywhere. Finally,
when $L_{u\phi}=0$, the last Bianchi identity reads
$\p_r (r L_{uu})=0$.  Thus the only non-vanishing term of $r L_{uu}$
is the constant one.

Accordingly, the equations of motions are solved in the following order:
\begin{itemize}
\item 4 main equations: $L_{rr}=0,\p_\nu(\sqrt{-g} F^{u
\nu})=0,L_{r\phi}=0,L_{ru}=0$,
\item 1 standard equation: $\p_\nu (\sqrt{-g }F^{\phi \nu})=0$,
\item 3 supplementary equations: $\p_\nu (\sqrt{-g }F^{r
    \nu})=0,L_{u\phi}=0,L_{uu}=0$,
\item 1 trivial equation: $L_{\phi\phi}=0$.
\end{itemize}

Starting with $L_{rr}=0$, we have $g_{rr}=0$,
$R_{rr}=2\dfrac{\p_r\beta}{r}$,
$T_{rr}=\dfrac{2}{r^2}(F_{r\phi})^2$. Hence
$\p_r\beta=\dfrac{1}{r}(F_{r\phi})^2$ and thus
$\beta=\beta_0(u,\phi)-\int_r^\infty dr'~~\dfrac{1}{r'}(F_{r\phi})^2$
with $\beta_0$ an integration constant. The fall-off condition
$\beta=o(r^{0})$ puts $\beta_0$ to zero and thus,
\begin{eqnarray}
\beta=-\int_r^\infty dr~~\dfrac{1}{r'}(F_{r\phi})^2\label{beta}.
\end{eqnarray}

Consider now the equation $\p_\nu(\sqrt{-g} F^{u \nu})=0$. Explicitly,
this equation reads $\p_r (r e^{2\beta} F^{u r})+ \p_\phi (r
e^{2\beta} F^{u \phi})=0$. Defining
\begin{equation}
m:=e^{2\beta}
F^{ur}=-e^{-2\beta}(F_{ur}-UF_{r\phi}),\label{eq:10}
\end{equation}
and using $e^{2\beta} F^{u\phi}=-\dfrac{1}{r^2}F_{r\phi}$, this
equation of motion is a first order differential equation for $m$,
\begin{equation}
  \p_r (r m)=\dfrac{\p_\phi F_{r\phi}}{r} \Longrightarrow
  m=\dfrac{-\lambda-\int_r^\infty
    dr'~~\dfrac{\p_\phi F_{r\phi}}{r'}}{r}, \label{E}
\end{equation}
with $\lambda(u,\phi)$ a constant of integration.

For $L_{r\phi}=0$, we have $g_{r\phi}=0$, $R_{r\phi}=-\p_{\phi
  r}\beta+\dfrac{\p_\phi\beta}{r}-r^2e^{-2\beta}\p_r\beta \p_r
U+\dfrac{3}{2}re^{-2\beta}\p_r U+\dfrac{r^2}{2}e^{-2\beta}\p_{rr}U$,
$T_{r\phi}=2F_{r\phi} m$. Defining
\begin{equation}
  n:=\dfrac{r^2}{2}e^{-2\beta}\p_rU,\label{eq:12}
\end{equation}
$R_{r\phi}=-\p_{\phi r}\beta+\dfrac{\p_\phi\beta}{r}+
\left(\partial_r+\dfrac{1}{r}\right)n$,
the equation is a first order differential equation for
$n$,
\begin{multline}
\p_rn+\dfrac{n}{r}=2F_{r\phi} m+\p_{r\phi}\beta-\dfrac{\p_\phi\beta}{r}\\
\Longrightarrow n=\dfrac{N
-2\int_r^\infty dr'~~r'(2F_{r\phi}m+\p_{r\phi}\beta-
\dfrac{\p_\phi\beta}{r})}{2r},\label{n}
\end{multline}
with $N(u,\phi)$ an integration constant. As a consequence of the
fall-off condition on $U$, we end up with
\begin{equation}
U=-\int_r^\infty dr'~~(\frac{2e^{2\beta}}{r'^2}n).\label{U}
\end{equation}

For $L_{ur}=0$, we have $G_{ru}=-\half
g_{ru}(R_{\phi\phi}g^{\phi\phi} + 2 R_{r\phi}g^{r\phi}
+R_{rr}g^{rr})$, $R_{\phi\phi}=r e^{-2\beta}( \p_rV + 2 \p_\phi
U)-2\p_{\phi\phi}\beta+r^2e^{-2\beta}\p_{\phi
  r}U-2(\p_\phi\beta)^2-\dfrac{e^{-4\beta}}{2}r^4(\p_r U)^2$ and $-2
r^2 (T_{ru}g^{ru} + T_{r\phi}g^{r\phi} + \half T_{rr}g^{rr})= 2 r^2
m^2$.
This gives
\begin{equation}
  \p_rV= 2 r e^{2\beta} m^2 + \frac{2 e^{2\beta}
    \p_{\phi\phi}\beta}{r} - r\p_{\phi r} U + \frac{2 e^{2\beta}
    (\p_\phi \beta)^2}{r} + \half e^{-2\beta}r^3(\p_r U)^2 - 2
  \p_\phi U,
\end{equation}
and
\begin{multline}
 V=\theta-\int_r^\infty dr' \Big(2 r
    e^{2\beta} m^2 + \frac{2 e^{2\beta} \p_{\phi\phi}\beta}{r} -
    r\p_{\phi r} U + \frac{2 e^{2\beta} (\p_\phi \beta)^2}{r} +\\+ \half
    e^{-2\beta}r^3(\p_r U)^2 - 2 \p_\phi U \Big),\label{V}
\end{multline}
with $\theta(u,\phi)$ a constant of integration.

For $j\geq 0$, we have
\begin{equation}
\p_r [r^i \ln^j r] = \left\{
{\renewcommand{\arraystretch}{2}
\begin{array}{ll}
\sum_{k=0}^j C_{ijk} r^{i-1} \ln^kr , &\mbox{$i \ne 0$} , \\
\displaystyle{{{j r^{-1}\ln^{j-1} r} }} , &\mbox{$i = 0$} ,
\end{array}} \right.
\end{equation}
\begin{equation}
\int r^i \ln^j r\  dr = \left\{
{\renewcommand{\arraystretch}{2}
\begin{array}{ll}
\sum_{k=0}^j D_{ijk} r^{i+1} \ln^kr , &\mbox{$i \ne -1$} , \\
\displaystyle{{\frac{\ln^{j+1} r}{j+1}}} , &\mbox{$i = -1$} ,
\end{array}} \right.
\end{equation}
for some coefficients $C_{ijk}$ and $D_{ijk}$, and up to constants for
the integrations. Consider then series ${\cal S}^{n}$ with elements of
the form
\begin{equation}
  \label{eq:1a}
  \sum_{i\leq -n,\, 0\leq j\leq -i-n} s_{ij}(u,\phi)r^i\ln^j r,
\end{equation}
with $n\geq 0$. These series satisfy
${\cal S}^{n+1}\subset {\cal S}^{n}$,
${\cal S}^{n}* {\cal S}^{m}\subset {\cal S}^{n+m}$,
$({\cal S}^{n})'\subset {\cal S}^{n+1}$. For integration however,
$\int dr\ {\cal S}^{n+1}\subset {\cal S}^{n}$, up to constants for
$n\neq 0$ and the divergent logarithmic term for $n=0$.

The ansatz \eqref{eq:8} belongs to ${\cal S}^{0}$, up to the divergent
logarithmic term proportional to $\alpha(u,\phi)$. This implies
$F_{r\phi}\in {\cal S}^{1}$ and $\beta$ in \eqref{beta}, because of
the absence of the constant, belongs to ${\cal S}^{2}$ with all
coefficients determined by the coefficients
$\alpha(u,\phi),A_{mn}(u,\phi)$ of \eqref{eq:8}.

In the same way, from $\eqref{E}$, it follows that $m\in {\cal S}^{1}$
with all coefficients determined by those of \eqref{eq:8} and the
integration function $\lambda$.

For $U$, we have in a first stage that $n$ belongs to ${\cal S}^{0}$
and is determined by the data in \eqref{eq:8} and the integration
constants $\lambda,N$. For $U$ itself, it follows from \eqref{U}, that it
belongs to ${\cal S}^{1}$, with no new integration constant because of
the assumed fall-off.

Finally, it follows from \eqref{V} that $V$ belongs to ${\cal S}^{0}$,
up to a logarithmic divergence, with coefficients determined by the
data in \eqref{eq:8} and the integration functions
$\alpha,\lambda,N,\theta$.

In summary, by integrating $m$ in $r$ in order to get $A_u$ and making
the $\alpha$ dependence explicit, we find that all main equations are
solved as
\begin{eqnarray}
  &&m=-\frac{\lambda}{r}-\frac{\alpha'}{r^2}
  +\sum\limits_{m=1}^\infty\sum\limits_{n=0}^{m}
\frac{m_{mn}(\ln \frac{r}{r_0})^n}{r^{m+2}},\\
  &&\beta=-\frac{\alpha^2}{2r^2}+\sum\limits_{m=1}^\infty
  \sum\limits_{n=0}^{m} \frac{\beta_{mn}(\ln \frac{r}{r_0})^n}{r^{m+2}},\\
  &&U=\frac{4\lambda \alpha \ln \frac{r}{r_0} + 2 \lambda \alpha -N}{2r^2}
  +\sum\limits_{m=1}^\infty\sum\limits_{n=0}^{m}
\frac{U_{mn}(\ln \frac{r}{r_0})^n}{r^{m+2}},\\
  &&A_u=-\lambda\ln {\frac{r}{r_0}}+A_u^0+\frac{
    \alpha'}{r}+\sum\limits_{m=1}^\infty
  \sum\limits_{n=0}^{m} \frac{B_{mn}(\ln \frac{r}{r_0})^n}{r^{m+1}},\\
  &&V=2 \lambda^2 \ln \frac{r}{r_0} + \theta
 +\frac{2\alpha \lambda' - 2\lambda \alpha'}{r},\\
  &&\hspace{1cm}+\sum\limits_{m=1}^\infty\sum\limits_{n=0}^{m}
\frac{V_{mn}(\ln
    \frac{r}{r_0})^n}{r^{m+1}},
\end{eqnarray}
where $m_{mn},\beta_{mn},U_{mn},V_{mn},B_{mn}$ are determined by $\alpha(u,\phi)$,
$A_{mn}(u,\phi)$, the integration constants
$\lambda(u,\phi),N(u,\phi)$ and their $\phi$
derivatives.

The standard equation determines the $u$ evolution of
$\alpha,A_\phi^0$ and $A_{mn}(u,\phi)$. Indeed,
$\p_\nu (\sqrt{-g}F^{\phi\nu})=\p_u (r e^{2\beta} F^{\phi u})+ \p_r (r
e^{2\beta} F^{\phi r})=0$.
Since
$e^{2\beta} F^{\phi r}=Um+\dfrac{1}{r^2}\left(F_{u\phi}+ V
  F_{r\phi}\right)$,
$e^{2\beta} F^{\phi u} = \dfrac{1}{r^2}F_{r\phi}$, we get
\begin{equation*}
\p_u  F_{r\phi}=-r^2\left(\partial_r+\dfrac{1}{r}\right)\left[U m
  +\dfrac{1}{r^2}\left(F_{u\phi}+V F_{r\phi}\right)\right],
\end{equation*}
which is a differential equation governing the $u$-dependence of
$F_{r\phi}$ and thus of $A_\phi$. In terms of coefficients, we get
\begin{equation}
\dot\alpha =-\lambda',\quad  \dot A_\phi^0 = -\lambda' + (A_u^0)',\quad
\dot A_{m n+1}=\frac{(2m+1)\dot A_{m n}+X_{m n}}{2(n+1)}, \label{AA}
\end{equation}
where $A_{m n+1}=0$ when $n=m$ and $X_{m n}$ is a linear
combination of $\alpha, A_\phi^0, A_{mn}$, integration functions
$\lambda, A_u^0, N, \theta$ and their $\phi$
derivative.

The first supplementary equation reads explicitly
$0=\p_\nu
(\sqrt{-g}F^{r\nu})=\p_u (r e^{2\beta} F^{r u})+
\p_\phi (r e^{2\beta} F^{r \phi})$. Since $e^{2\beta} F^{ru}=-
m=\frac{\lambda}{r} + O(r^{-2})$ and $e^{2\beta} F^{r\phi}=\left[U
  m-\dfrac{1}{r^2}\left(F_{u\phi}+
\dfrac{V}{r}F_{r\phi}\right)\right]=O(r^{-2})$,
$\lim_{r\to\infty} \p_\nu (\sqrt{-g}F^{r\nu})=0$ implies
$\dot \lambda = 0$ so that $\lambda = \lambda(\phi)$.
The first of equation \eqref{AA} then implies
$\alpha=-\omega(\phi)-u\lambda'$.

For the second supplementary equation, $L_{u\phi}=0$, we have
$L_{u\phi}=\dfrac{1}{2r}(\theta'-\dot N)+O(r^{-2})$. Hence,
$\lim_{r\to\infty}(rL_{u \phi})=0$ implies $\dot N=\theta'$.

For the last supplementary equation $L_{uu}=0$, we have
$L_{uu}=\dfrac{\dot \theta}{r}+O(r^{-2})$.
$\lim_{r\to\infty}(rL_{uu})=0$ then implies $\partial_u \theta=0$ and thus
$\theta=\theta(\phi)$ and then also that $N=\chi(\phi)+u \theta'$.


\begin{thebibliography}{10}

\bibitem{Bondi:1962px}
H.~Bondi, M.~G. van~der Burg, and A.~W. Metzner, ``{Gravitational waves In
  general relativity. 7. {W}aves from axi-symmetric isolated systems},''
{\em Proc.\ Roy.\ Soc.\ Lond.\ A} {\bfseries 269} (1962) 21.

\bibitem{Sachs1962a}
R.~K. Sachs, ``Gravitational waves in general relativity. 8. {W}aves in
  asymptotically flat space-time,''
{\em Proc.\ Roy.\ Soc.\ Lond.\ A} {\bfseries 270} (1962) 103.

\bibitem{Newman:1963yy}
E.~Newman, L.~Tamubrino, and T.~Unti, ``Empty space generalization of the
  {S}chwarzschild metric,''
{\em J. Math. Phys.} {\bfseries 4} (1963) 915.

\bibitem{Burg1969}
M.~G.~J. van~der Burg, ``{Gravitational Waves in General Relativity. 10.
  Asymptotic Expansions for the Einstein-Maxwell Field},''
  \href{http://dx.doi.org/10.1098/rspa.1969.0072}{{\em Proc.\ Roy.\ Soc.\
  Lond.\ A} {\bfseries 310} (1969) 221--230}.

\bibitem{Exton1969}
A.~Exton, E.~Newman, and R.~Penrose, ``{Conserved quantities in the
  Einstein-Maxwell theory},'' {\em J.Math.Phys.} {\bfseries 10} (1969)
  1566--1570.

\bibitem{Winicour:1985pi}
J.~Winicour, ``Logarithmic asymptotic flatness,'' {\em Foundations of Physics}
  {\bfseries 15} no.~5, (05, 1985) 605--616.
  \url{http://dx.doi.org/10.1007/BF01882485}.

\bibitem{Chrusciel:1993hx}
P.~T. Chrusciel, M.~A.~H. MacCallum, and D.~B. Singleton, ``{Gravitational
  waves in general relativity. 14. Bondi expansions and the polyhomogeneity of
  Scri},'' {\em Proc.\ Roy.\ Soc.\ Lond.\ A} {\bfseries 436} (1992) 299--316,
\href{http://arxiv.org/abs/gr-qc/9305021}{{\ttfamily arXiv:gr-qc/9305021}}.

\bibitem{ValienteKroon:1998vn}
J.~A. Valiente~Kroon, ``{Logarithmic Newman-Penrose constants for arbitrary
  polyhomogeneous space-times},''
  \href{http://dx.doi.org/10.1088/0264-9381/16/5/314}{{\em Class.Quant.Grav.}
  {\bfseries 16} (1999) 1653--1665},
\href{http://arxiv.org/abs/gr-qc/9812004}{{\ttfamily arXiv:gr-qc/9812004
  [gr-qc]}}.

\bibitem{Wald:1999wa}
R.~M. Wald and A.~Zoupas, ``A general definition of conserved quantities in
  general relativity and other theories of gravity,'' {\em Phys. Rev.}
  {\bfseries D61} (2000) 084027,
\href{http://arxiv.org/abs/gr-qc/9911095}{{\ttfamily gr-qc/9911095}}.

\bibitem{Barnich:2009se}
G.~Barnich and C.~Troessaert, ``{Symmetries of asymptotically flat 4
  dimensional spacetimes at null infinity revisited},''
  \href{http://dx.doi.org/10.1103/PhysRevLett.105.111103}{{\em Phys.Rev.Lett.}
  {\bfseries 105} (2010) 111103},
\href{http://arxiv.org/abs/0909.2617}{{\ttfamily arXiv:0909.2617 [gr-qc]}}.

\bibitem{Barnich:2010eb}
G.~Barnich and C.~Troessaert, ``{Aspects of the BMS/CFT correspondence},''
  \href{http://dx.doi.org/10.1007/JHEP05(2010)062}{{\em JHEP} {\bfseries 05}
  (2010) 062},
\href{http://arxiv.org/abs/1001.1541}{{\ttfamily arXiv:1001.1541 [hep-th]}}.

\bibitem{Barnich:2011mi}
G.~Barnich and C.~Troessaert, ``{BMS charge algebra},''
  \href{http://dx.doi.org/10.1007/JHEP12(2011)105}{{\em JHEP} {\bfseries 1112}
  (2011) 105},
\href{http://arxiv.org/abs/1106.0213}{{\ttfamily arXiv:1106.0213 [hep-th]}}.

\bibitem{Barnich:2013axa}
G.~Barnich and C.~Troessaert, ``{Comments on holographic current algebras and
  asymptotically flat four dimensional spacetimes at null infinity},''
  \href{http://dx.doi.org/10.1007/JHEP11(2013)003}{{\em JHEP} {\bfseries 1311}
  (2013) 003},
\href{http://arxiv.org/abs/1309.0794}{{\ttfamily arXiv:1309.0794 [hep-th]}}.

\bibitem{Ashtekar1997}
A.~Ashtekar, J.~Bicak, and B.~G. Schmidt, ``Asymptotic structure of
  symmetry-reduced general relativity,'' {\em Phys. Rev. D} {\bfseries 55}
  no.~2, (Jan, 1997) 669--686.

\bibitem{Compere:2014cna}
G.~Comp\`ere, L.~Donnay, P.-H. Lambert, and W.~Schulgin, ``{Liouville theory
  beyond the cosmological horizon},''
  \href{http://dx.doi.org/10.1007/JHEP03(2015)158}{{\em JHEP} {\bfseries 03}
  (2015) 158},
\href{http://arxiv.org/abs/1411.7873}{{\ttfamily arXiv:1411.7873 [hep-th]}}.

\bibitem{Barnich:2006avcorr}
G.~Barnich and G.~Comp{\`e}re, ``Classical central extension for asymptotic
  symmetries at null infinity in three spacetime dimensions,'' {\em Class.
  Quant. Grav.} {\bfseries 24} (2007) F15,
  \href{http://arxiv.org/abs/gr-qc/0610130}{{\ttfamily gr-qc/0610130}}.
Corrigendum: ibid 24 (2007) 3139.

\bibitem{Brown:1986nw}
J.~D. Brown and M.~Henneaux, ``Central charges in the canonical realization of
  asymptotic symmetries: An example from three-dimensional gravity,'' {\em
  Commun. Math. Phys.} {\bfseries 104} (1986) 207.

\bibitem{Deser:1983tn}
S.~Deser, R.~Jackiw, and G.~'t~Hooft, ``Three-dimensional {E}instein gravity:
  Dynamics of flat space,''
{\em Ann. Phys.} {\bfseries 152} (1984) 220.

\bibitem{Ezawa:1992nk}
K.~Ezawa, ``{Transition amplitude in (2+1)-dimensional Chern-Simons gravity on
  a torus},'' \href{http://dx.doi.org/10.1142/S0217751X94001898}{{\em
  Int.J.Mod.Phys.} {\bfseries A9} (1994) 4727--4746},
\href{http://arxiv.org/abs/hep-th/9305170}{{\ttfamily arXiv:hep-th/9305170
  [hep-th]}}.

\bibitem{Cornalba:2002fi}
L.~Cornalba and M.~S. Costa, ``{A New cosmological scenario in string
  theory},'' \href{http://dx.doi.org/10.1103/PhysRevD.66.066001}{{\em
  Phys.Rev.} {\bfseries D66} (2002) 066001},
\href{http://arxiv.org/abs/hep-th/0203031}{{\ttfamily arXiv:hep-th/0203031
  [hep-th]}}.

\bibitem{Cornalba:2003kd}
L.~Cornalba and M.~S. Costa, ``{Time dependent orbifolds and string
  cosmology},'' \href{http://dx.doi.org/10.1002/prop.200310123}{{\em
  Fortsch.Phys.} {\bfseries 52} (2004) 145--199},
\href{http://arxiv.org/abs/hep-th/0310099}{{\ttfamily arXiv:hep-th/0310099
  [hep-th]}}.

\bibitem{Barnich:2012aw}
G.~Barnich, A.~Gomberoff, and H.~A. Gonzalez, ``{Flat limit of three
  dimensional asymptotically anti-de Sitter spacetimes},''
  \href{http://dx.doi.org/10.1103/PhysRevD.86.024020}{{\em Phys.Rev.}
  {\bfseries D86} (2012) 024020},
\href{http://arxiv.org/abs/1204.3288}{{\ttfamily arXiv:1204.3288 [gr-qc]}}.

\bibitem{Martinez:1999qi}
C.~Martinez, C.~Teitelboim, and J.~Zanelli, ``{Charged rotating black hole in
  three space-time dimensions},''
  \href{http://dx.doi.org/10.1103/PhysRevD.61.104013}{{\em Phys.Rev.}
  {\bfseries D61} (2000) 104013},
\href{http://arxiv.org/abs/hep-th/9912259}{{\ttfamily arXiv:hep-th/9912259
  [hep-th]}}.

\bibitem{Sachs1962}
R.~K. Sachs, ``Asymptotic symmetries in gravitational theory,'' {\em Phys.\
  Rev.} {\bfseries 128} (1962) 2851--2864.

\bibitem{Barnich:2013sxa}
G.~Barnich and P.-H. Lambert, ``{Einstein-Yang-Mills theory: Asymptotic
  symmetries},'' \href{http://dx.doi.org/10.1103/PhysRevD.88.103006}{{\em
  Phys.Rev.} {\bfseries D88} (2013) 103006},
\href{http://arxiv.org/abs/1310.2698}{{\ttfamily arXiv:1310.2698 [hep-th]}}.

\bibitem{Tamburino1966}
L.~A. Tamburino and J.~H. Winicour, ``Gravitational fields in finite and
  conformal {B}ondi frames,'' {\em Phys. Rev.} {\bfseries 150} (1966) 1039.

\bibitem{Anderson:1996sc}
I.~M. Anderson and C.~G. Torre, ``Asymptotic conservation laws in field
  theory,'' {\em Phys. Rev. Lett.} {\bfseries 77} (1996) 4109--4113,
\href{http://arxiv.org/abs/hep-th/9608008}{{\ttfamily hep-th/9608008}}.

\bibitem{Barnich:2001jy}
G.~Barnich and F.~Brandt, ``Covariant theory of asymptotic symmetries,
  conservation laws and central charges,'' {\em Nucl. Phys.} {\bfseries B633}
  (2002) 3--82,
\href{http://arXiv.org/abs/hep-th/0111246}{{\ttfamily hep-th/0111246}}.

\bibitem{Barnich:2003xg}
G.~Barnich, ``Boundary charges in gauge theories: {U}sing {S}tokes theorem in
  the bulk,'' {\em Class. Quant. Grav.} {\bfseries 20} (2003) 3685--3698,
\href{http://arxiv.org/abs/hep-th/0301039}{{\ttfamily hep-th/0301039}}.

\bibitem{Barnich:2007bf}
G.~Barnich and G.~Comp\`{e}re, ``{Surface charge algebra in gauge theories and
  thermodynamic integrability},''
  \href{http://dx.doi.org/10.1063/1.2889721}{{\em J. Math. Phys.} {\bfseries
  49} (2008) 042901},
\href{http://arxiv.org/abs/0708.2378}{{\ttfamily arXiv:0708.2378 [gr-qc]}}.

\bibitem{Barnich:2005kq}
G.~Barnich and G.~Comp{\`e}re, ``Conserved charges and thermodynamics of the
  spinning {G}oedel black hole,'' {\em Phys. Rev. Lett.} {\bfseries 95} (2005)
  031302,
\href{http://arxiv.org/abs/hep-th/0501102}{{\ttfamily hep-th/0501102}}.

\bibitem{Hartnoll:2008vx}
S.~A. Hartnoll, C.~P. Herzog, and G.~T. Horowitz, ``{Building a Holographic
  Superconductor},''
  \href{http://dx.doi.org/10.1103/PhysRevLett.101.031601}{{\em Phys. Rev.
  Lett.} {\bfseries 101} (2008) 031601},
\href{http://arxiv.org/abs/0803.3295}{{\ttfamily arXiv:0803.3295 [hep-th]}}.

\bibitem{Kachru:2008yh}
S.~Kachru, X.~Liu, and M.~Mulligan, ``{Gravity duals of Lifshitz-like fixed
  points},'' \href{http://dx.doi.org/10.1103/PhysRevD.78.106005}{{\em Phys.
  Rev.} {\bfseries D78} (2008) 106005},
\href{http://arxiv.org/abs/0808.1725}{{\ttfamily arXiv:0808.1725 [hep-th]}}.

\bibitem{Balasubramanian:2008dm}
K.~Balasubramanian and J.~McGreevy, ``{Gravity duals for non-relativistic
  CFTs},'' \href{http://dx.doi.org/10.1103/PhysRevLett.101.061601}{{\em Phys.
  Rev. Lett.} {\bfseries 101} (2008) 061601},
\href{http://arxiv.org/abs/0804.4053}{{\ttfamily arXiv:0804.4053 [hep-th]}}.

\bibitem{Maity:2009zz}
D.~Maity, S.~Sarkar, N.~Sircar, B.~Sathiapalan, and R.~Shankar, ``{Properties
  of CFTs dual to Charged BTZ black-hole},''
  \href{http://dx.doi.org/10.1016/j.nuclphysb.2010.06.012}{{\em Nucl. Phys.}
  {\bfseries B839} (2010) 526--551},
\href{http://arxiv.org/abs/0909.4051}{{\ttfamily arXiv:0909.4051 [hep-th]}}.

\bibitem{Ren:2010ha}
J.~Ren, ``{One-dimensional holographic superconductor from AdS$\_3$/CFT$\_2$
  correspondence},'' \href{http://dx.doi.org/10.1007/JHEP11(2010)055}{{\em
  JHEP} {\bfseries 11} (2010) 055},
\href{http://arxiv.org/abs/1008.3904}{{\ttfamily arXiv:1008.3904 [hep-th]}}.

\bibitem{Jensen:2010em}
K.~Jensen, ``{Chiral anomalies and AdS/CMT in two dimensions},''
  \href{http://dx.doi.org/10.1007/JHEP01(2011)109}{{\em JHEP} {\bfseries 01}
  (2011) 109},
\href{http://arxiv.org/abs/1012.4831}{{\ttfamily arXiv:1012.4831 [hep-th]}}.

\bibitem{Bagchi:2009pe}
A.~Bagchi, R.~Gopakumar, I.~Mandal, and A.~Miwa, ``{GCA in 2d},''
  \href{http://dx.doi.org/10.1007/JHEP08(2010)004}{{\em JHEP} {\bfseries 08}
  (2010) 004},
\href{http://arxiv.org/abs/0912.1090}{{\ttfamily arXiv:0912.1090 [hep-th]}}.

\bibitem{Bagchi:2010zz}
A.~Bagchi, ``{Correspondence between Asymptotically Flat Spacetimes and
  Nonrelativistic Conformal Field Theories},''
\href{http://dx.doi.org/10.1103/PhysRevLett.105.171601}{{\em Phys.Rev.Lett.}
  {\bfseries 105} (2010) 171601}.

\bibitem{Ashtekar:1987tt}
A.~Ashtekar, ``{Asymptotic Quantization: Based on 1984 Naples Lectures},''.
  Naples, Italy: Bibliopolis (1987) 107 p. (Monographs and textbooks in
  physical science, 2).

\bibitem{Barnich:2011ct}
G.~Barnich and C.~Troessaert, ``{Supertranslations call for superrotations},''
  {\em {PoS}} {\bfseries CNCFG2010} (2010) 010,
\href{http://arxiv.org/abs/1102.4632}{{\ttfamily arXiv:1102.4632 [gr-qc]}}.

\bibitem{Barnich:2012xq}
G.~Barnich, ``{Entropy of three-dimensional asymptotically flat cosmological
  solutions},'' \href{http://dx.doi.org/10.1007/JHEP10(2012)095}{{\em JHEP}
  {\bfseries 1210} (2012) 095},
\href{http://arxiv.org/abs/1208.4371}{{\ttfamily arXiv:1208.4371 [hep-th]}}.

\bibitem{Bagchi:2012xr}
A.~Bagchi, S.~Detournay, R.~Fareghbal, and J.~Simon, ``{Holography of 3d Flat
  Cosmological Horizons},''
  \href{http://dx.doi.org/10.1103/PhysRevLett.110.141302}{{\em Phys.Rev.Lett.}
  {\bfseries 110} (2013) 141302},
\href{http://arxiv.org/abs/1208.4372}{{\ttfamily arXiv:1208.4372 [hep-th]}}.

\bibitem{Strominger:1998eq}
A.~Strominger, ``Black hole entropy from near-horizon microstates,'' {\em JHEP}
  {\bfseries 02} (1998) 009,
\href{http://arxiv.org/abs/arXiv:hep-th/9712251}{{\ttfamily
  arXiv:hep-th/9712251}}.

\bibitem{ValienteKroon:2000jy}
J.~A. Valiente-Kroon, ``{On the existence and convergence of polyhomogeneous
  expansions of zero rest mass fields},''
  \href{http://dx.doi.org/10.1088/0264-9381/17/21/302}{{\em Class.Quant.Grav.}
  {\bfseries 17} (2000) 4365--4376},
\href{http://arxiv.org/abs/gr-qc/0005087}{{\ttfamily arXiv:gr-qc/0005087
  [gr-qc]}}.

\end{thebibliography}


\section*{References}
\addcontentsline{toc}{section}{References}

\renewcommand{\section}[2]{}%

\def\cprime{$'$}
\providecommand{\href}[2]{#2}\begingroup\raggedright\endgroup

\end{document}